\documentclass[12pt,preprint]{aastex}

\def\Msun{\ensuremath{\rm M_{\odot}}}
\def\kms{km~s\ensuremath{^{-1}}}
\def\wn{cm\ensuremath{^{-1}}}
\def\col{mol~cm\ensuremath{^{-2}}}

\def\jchphys{JChPhys}

\shorttitle{OH in Disks Around Intermediate-Mass Stars}
\shortauthors{Mandell, Mumma, Blake, Bonev, Villanueva, \& Salyk}

\begin{document}
\title{Discovery of OH in Circumstellar Disks Around Young Intermediate-Mass Stars} 
\author{Avi M. Mandell\altaffilmark{1,5}, Michael J. Mumma\altaffilmark{1}, Goeffrey A. Blake\altaffilmark{2}, Boncho P. Bonev\altaffilmark{1,3}, Geronimo L. Villanueva\altaffilmark{1,5}, \& Colette Salyk\altaffilmark{2}} 
\altaffiltext{1}{NASA Goddard Space Flight Center, Code 693, Greenbelt, MD 20771, USA}
\altaffiltext{2}{California Institute of Technology, Division of Geological and Planetary Sciences, MS 150-21, Pasadena, CA 91125, USA}
\altaffiltext{3}{Catholic University of America, Department of Physics, Washington, DC 20064, USA}
\altaffiltext{4}{Corresponding Email:  Avi.Mandell@nasa.gov}
\altaffiltext{5}{NPP Postdoctoral Fellow}

\begin{abstract}
We detect emission from multiple low-excitation ro-vibrational transitions of OH from the two Herbig Ae stars AB Aurigae and MWC 758 in the $3.0 - 3.7 \micron$ wavelength range (L-band), using the NIRSPEC instrument on Keck II.  The inner radius for the emitting region in both stars is close to 1 AU. We compare an optically thin LTE model and a thin-wedge fluorescence model, finding rotational temperatures of $650-800$ K and OH abundances of $10^{42}-10^{45}$ molecules for the two stars. Comparisons with current chemical models support the fluorescence excitation model for AB Aurigae and possibly MWC 758, but further observations and detailed modeling are necessary to improve constraints on OH emission in different disk environments.  
\end{abstract}


\section{Introduction}
We report the first detection of warm hydroxyl (OH) emission from the circumstellar disks of two young intermediate-mass stars. AB Aurigae (HD 31293) has a mass of 2.4 \Msun\ and an age of 1-3 Myr, while MWC 758 (HD 36112) has a mass of 2.0 \Msun\ and an age of 2-5 Myr \citep{vandenancker98}.  Each disk is seen at low inclination, implying that our observations probe the surface layer and warm inner region ($R < 20$ AU) of the circumstellar material. 

The detection of OH provides a new window into the gas chemistry of these disks. OH has a central role in many chemical reactions involving the formation and destruction of H$_2$, CO, and H$_2$O, and the detection of warm hydroxyl opens up an excellent new diagnostic of the thermal and radiative environment in the planet formation zone of circumstellar disks. We achieve very high S/N results using improved data reduction algorithms, allowing us to sensitively test different excitation models. These results, combined with similar data for an expanding suite of molecular tracers such as CO and H$_2$O, will significantly improve our ability to constrain the disk chemistry driven by intense radiation from the central star.

\section{Observations and Data Reduction}
\label{obs}
Spectra were acquired at high resolving power ($\lambda/\Delta\lambda \sim $ 27,000) using the Near-IR Spectrograph (NIRSPEC) on the 10-m Keck II telescope \citep{mclean98}.  P-branch transitions ($J''=2.5$ to $J''=17.5$) in the OH (1-0) band were sampled on observing runs in April, October, and December 2006. Each science target was observed on all three runs; the comparison star HR1620 was also observed as close in time to the two science targets as possible (on the December run a second comparison star (HR2714) was observed as an additional check on the reduction methods).  The telescope was nodded 12 arcsec in an ABBA sequence, with 60 seconds integration per beam.  

\begin{figure}
\plotone{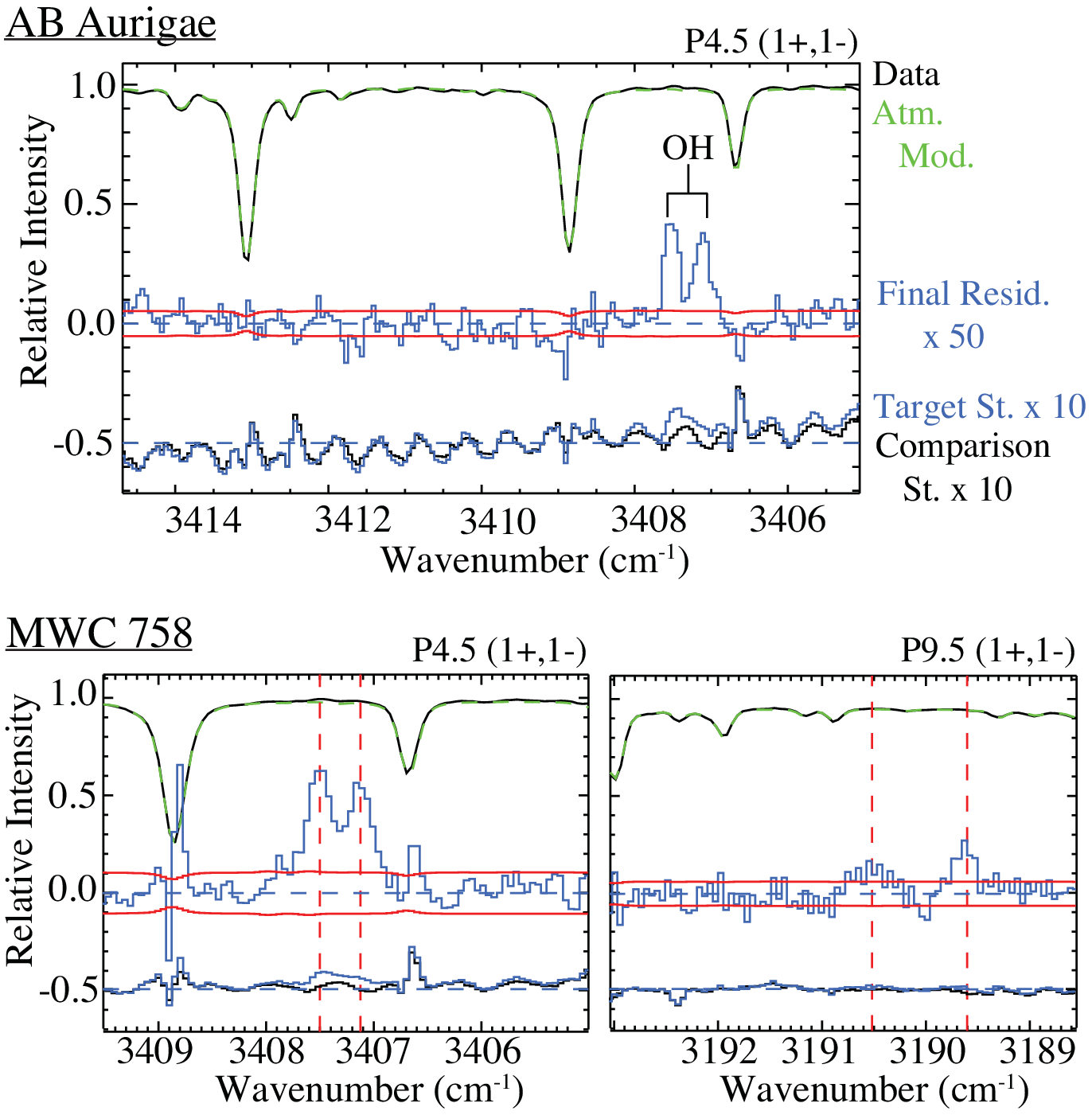}
\caption{\label{example} Examples of line detections for both stars. The upper section of each frame shows the stellar spectrum with terrestrial lines, as measured; a synthetic atmospheric spectrum is overlaid for comparison.  At the bottom of each frame are individual residuals for the science target and the comparison star after removal of a model atmosphere.  The middle residuals are the difference of the science and comparison star residuals bounded by 3-$\sigma$ photon-noise uncertainty limits.  A S/N $\sim$ 2000 was achieved in all spectral regions without severe telluric absorption.}
\end{figure}

Spectra were extracted and processed for each echelle order in each ABBA set using custom data reduction algorithms \citep{mandell07b,villanueva08?}; in this Letter we provide a summary of these procedures, with further details to be presented in a follow-up paper. The terrestrial spectral transmittance was synthesized for each ABBA set using the GENLN2 atmospheric code \citep{edwards92,hewagama03} and subtracted from the stellar spectrum; the difference removed the stellar continuum and compensated for changing airmass and telluric water vapor.  We co-added the residuals for all sets of each star, and the mean residuals of the science and comparison stars were then differenced, removing remnant fringes and other instrumental artifacts.  This process achieved photon-noise-limited results corresponding to S/N $\sim 2000$ on the original stellar continuum (see Figure \ref{example}).  Flux calibration was performed by comparing the observed continuum flux near each transition to the predicted flux based on the L-band stellar magnitude and temperature. Magnitudes were taken from \citet{hillenbrand92} (AB Aurigae) and \citet{malfait98} (MWC 758); for both stars, we used temperatures derived from fits to the NIR spectral energy distribution by \citet{natta01}.

The measured shapes of detected lines are asymmetric to varying degrees because they remain convolved with the atmospheric transmittance function; other unknown factors may also contribute.  The strongest OH doublet detected (P4.5 1+,1-) is not severely affected by local terrestrial lines and the line shapes are adequately reproduced by a Gaussian emission kernel convolved with an instrumental line shape function (Gaussian, with parameters determined from measured atmospheric features).  We solved for the ``best-fit" velocity width and radial velocity offset for the emission kernel, and assumed these values for other OH lines.  We then corrected the line areas measured for other OH doublets by fitting the data and measured errors to the scaled Gaussian line shape model convolved with the atmospheric transmittance in the local spectral region and the instrumental line shape; the uncertainty for each line was calculated using the  standard errors for the derived model parameters of the fit.  

\section{Results}

\begin{figure}
\plotone{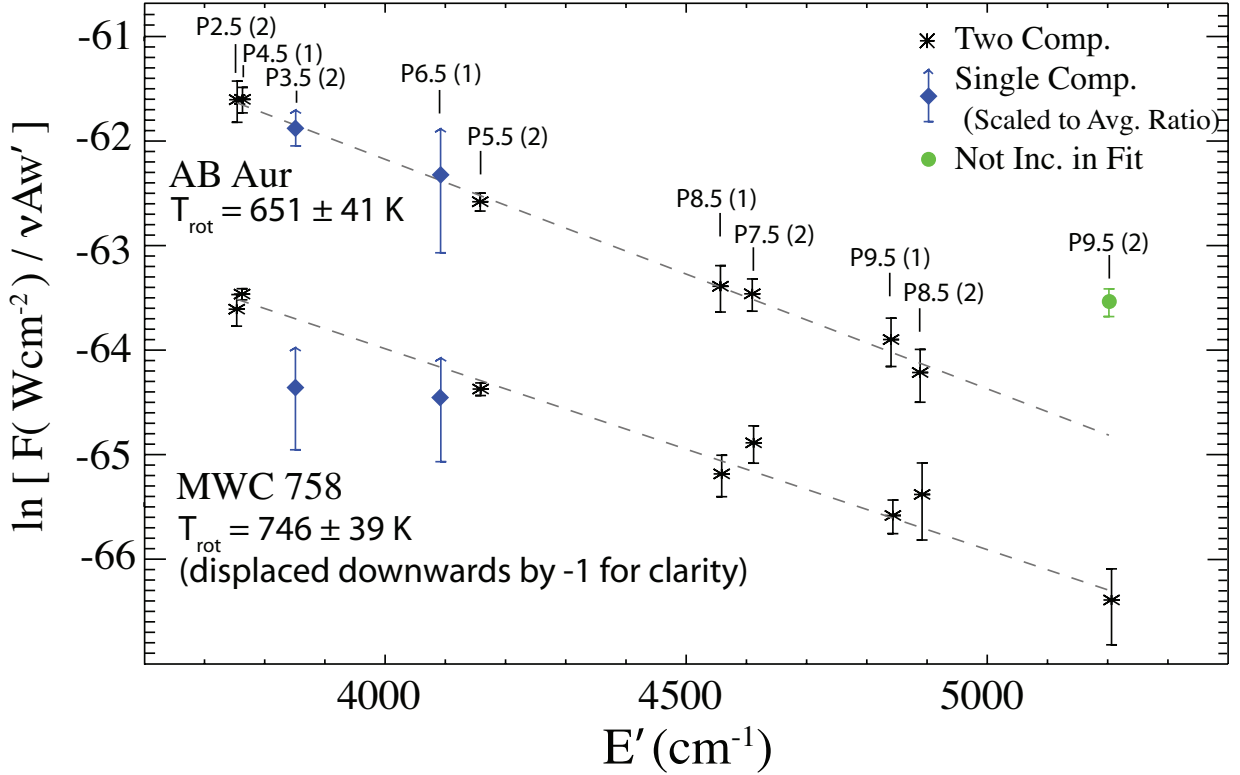}
\caption{\label{rotdiag} Rotation diagrams of detected transitions for AB Aurigae and MWC 758, with both doublet components averaged to minimize uncertainties. We detected all transitions with $J'' \le 9.5$ having transmittance above the noise limit; for two transitions (P3.5 (2), P6.5 (1)), only one component was measurable due to the atmospheric transmittance and therefore only the lower limit is secure. The P9.5 (2+, 2-) transition was not included in the fit for AB Aurigae; the inconsistently high flux may be due to a blend or an instrumental effect.}
\end{figure}

For both stars, either one or both $\Lambda$-doublet components were detected for all available OH $v=1-0$ ro-vibrational transitions ranging from $J'' = 2.5$ up to $J'' = 9.5$ and excitation energies up to 1600 \wn. Scaled line fluxes for both stars are shown in Figure \ref{rotdiag}; individual fluxes and uncertainties for each line will be tabulated in a follow-up paper. Lines were detected in all three sets of observations, and for each star the detected lines were found to have a common radial velocity shift (within 3 \kms) relative to the barycenter. Radial velocities derived from fits to the strongest and clearest doublet (P4.5 1+,1-) are $16.9\pm1.1$ and $15.7\pm0.7$ for AB Aurigae and MWC 758 respectively; these values compare well with radial velocities derived from sub-mm CO observations \citep{dent05}.  At NIRSPEC resolution (13 \kms) the line shapes of the strongest doublet (P4.5) are only marginally resolved; the FWHM for the P4.5 components (iteratively deconvolved from the instrumental profile as described in \S\ref{obs}) is $16.2\pm2.8$ \kms\ and $22.1\pm1.5$ \kms\ for AB Aurigae and MWC 758, respectively.  

The measured line fluxes can be related to an excitation temperature $T_{ex}$ and total OH abundance $N_{tot}$ using models for the excitation physics and radiative transfer. The simplest model assumes a collisionally-dominated optically thin medium in which the vibrational and rotational distributions for OH are described by a single temperature (LTE). However, if the emitting OH is located in a region of the disk atmosphere where the timescale for collisional excitation is longer than the timescale for radiative excitation of ground-state OH, rotational state distributions in the $v=1$ vibrational level is instead governed by both the ground state populations and the incident radiation field. We consider a simple collisional excitation model in \S \ref{lte} and a radiative excitation model in \S \ref{flo}; non-LTE collisional effects are deferred for future consideration.

\subsection{Collisional Excitation Model}
\label{lte}
When collisions dominate the excitation process, the emission intensity of ro-vibrational transitions varies linearly with the number of molecules and logarithmically with the excitation temperature \citep{herzberg50}. A rotation diagram, graphing the mean flux for each doublet pair (normalized by $\nu A\, w'$) against the energy of the excited level ($J',v=1$), is shown in Figure \ref{rotdiag}; transition parameters are taken from the HITRAN database \citep{rothman05}.  A linear fit to the rotation diagram for each star yields the excitation temperature ($T_{ex}$),and we use the $E'=0$ intercept from each fit multiplied by the total partition function for the derived temperatures (from HITRAN) and scaling by the distance to each star \citep{vandenancker98} to find the total number of OH molecules. Derived values are listed in Table \ref{table}.

An inner radius ($R_{in}$) for the OH emitting region can be found if the emission lines are broadened primarily by the projected radial velocities of material in Keplerian orbits around the central star. We assume a maximum orbital velocity of 2$\times$HWHM of the P4.5 lines, after removing the contribution from thermal broadening ($\sim$0.8 \kms\ for OH at 700K). However, the derived inner radius varies with the local disk inclination ($i$) as sin$^{2}(i)$, and published inclinations for circumstellar disks from NIR interferometry can vary substantially. For MWC 758, values from Eisner et al. (2004, hereafter E04) and Isella et al. (2006, hereafter I06) are similar, but for AB Aurigae values of sin($i$) derived by E04 and I06 differ by more than a factor of 2. Values for $R_{in}$ calculated using both sets of inclinations are given in Table \ref{table}; an outer radius for the emitting region is not constrained by the line shapes due to the low disk inclinations and limited spectral resolution.

\subsection{Radiative Excitation Model}
\label{flo}
Fluorescent OH emission can occur in diffuse astrophysical environments that are strongly irradiated, but the excitation details depend critically on the local UV and IR radiation field. OH can be excited to the $v = 1$ vibrational level (in the ground electronic state) both through ro-vibrational transitions in the near-IR as well as electronic transitions in the UV followed by cascades to the ground electronic state.  OH fluorescence has been studied extensively in comets \citep{schleicher82, weaver84, schleicher88}, and UV fluorescence of CO has recently been explored in disks \citep{brittain03, brittain07}. 

\begin{figure}
\plotone{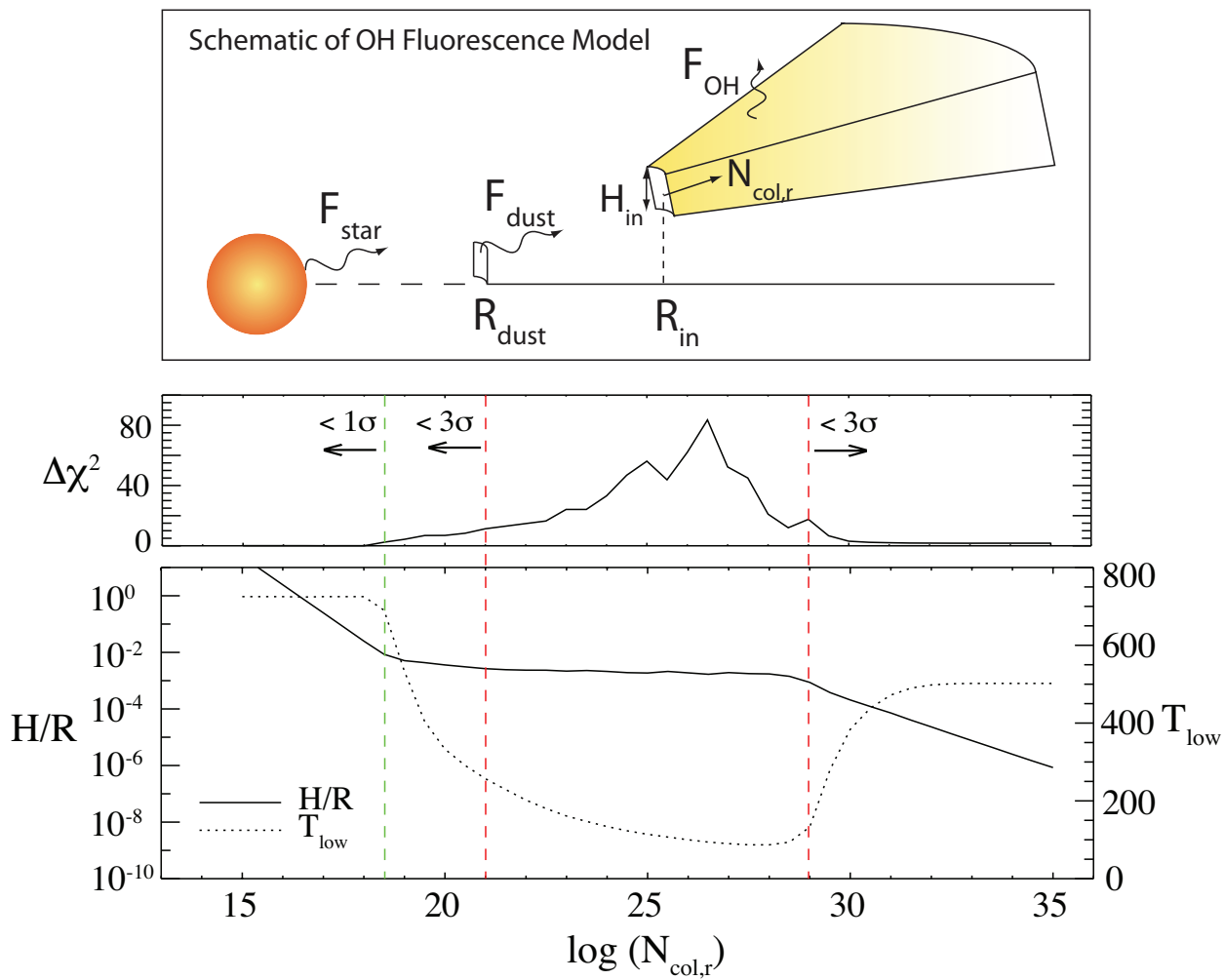}
\caption{\label{fluor} Predictions of a thin wedge fluorescence model for AB Aurigae. For a specific absorbing column density, a lower-state rotational temperature (T$_{low}$) and wedge thickness ($H/R$) are derived to match the observed component-averaged fluxes for each transition; the $\Delta\chi^2$ value (compared to the minimum value of 1.52) and standard confidence intervals for the fit are also shown.}
\end{figure}

We assess the effectiveness of OH fluorescence through a fit to a simple thin wedge model, with free parameters being the scaled thickness of the wedge ($H/R$), the column density in the absorbing (radial) direction at the inner boundary ($N_{col,r}$), and the lower-state rotational temperature ($T_{low}$) (Figure \ref{fluor}).  Integrating over the wedge geometry conserves the total number of available photons with distance from the star, and the use of an absorbing column density removes the need for a radial density profile. For each $N_{col,r}$, an optical depth and equivalent width is calculated for each transition and multiplied by the radiative flux at the minimum orbital distance ($R_{in}$, as defined in \S \ref{lte}) to find the total absorbed flux.  Model parameters ($T_{low}$ and $H/R$) are constrained by comparing predicted and measured fluxes for each transition; the $\chi^2$ value for the fit is used to evaluate the effects of variations in optical depth for different transitions. Multiplying $N_{col,r}$ by $H$ at the inner boundary gives the total number of molecules $N_{tot}$.

The fluorescent emission per molecule is calculated by summing the pump rates for transitions from all levels that contribute to state $J'$ using the Einstein B coefficients and then calculating the fraction of electrons that decay into state $J''$ using branching ratios based on the Einstein A coefficients \citep{disanti06}. For UV fluorescence, we consider pumping into the three lowest vibrational levels in the first electronic state ($A^2 \Sigma$) and decay into $v'' = 1$ in the ground state ($X^2 \Pi$) using rotational transition parameters derived from published vibrational band frequencies and transition probabilities \citep{schleicher88}. UV fluxes in the appropriate spectral region are taken from IUE spectra for both AB Aurigae and MWC 758 \citep{rodriguez99}; our approach does not compensate for flux variations on small spectral scales that cause UV fluorescence variations in comets \citep{swings41} but is adequate for this study. For IR fluorescence, we calculate pumping and decay rates using parameters from HITRAN. We adopt a basic 2-component model for the IR radiation field, comprised of thermal emission from both the star and an annular wall of ``hot dust" surrounding it \citep{natta01}; comparisons of our measured L-band fluxes for AB Aurigae and MWC 758 with those of standard stars of the same spectral type suggest the ``wall" emits approximately 10 times more L-band flux than the star alone. 

The location of the NIR-emitting dust region for both stars is uncertain. SEDs of HAe/Be stars are fit best by dust with temperatures of $\sim$1500K (the approximate sublimation temperature for ``astronomical silicate" dust) for the inner rim \citep{natta01}. If we assume the dust has an equilibrium temperature determined solely by the stellar radiation field, the dust rim would be located at a radius of 0.48 AU for AB Aurigae and 0.33 AU for MWC 758 (using stellar luminosities and stellar radii tabulated by \citet{vandenancker98}). These values are consistent with the analysis of NIR interferometric data by I06; however, earlier work by E04 found significantly smaller inner dust radii for the two stars. For completeness, we ran models using inner dust radii and disk inclinations (important for defining $R_{in}$) from both studies; model predictions are tabulated as a range of values for each star in Table \ref{table}.

Predictions of our thin wedge model for various parameters (slab thickness ($H/R$), lower-state temperature ($T_{low}$), and $\chi^2$), iterated over different values of the absorbing OH column density, are shown for AB Aurigae in Figure \ref{fluor}. A $\Delta\chi^2$ of less than 9 is predicted by either optically thin or optically thick absorbing column; however, optically thin models have $H/R > 10^{-3}$ while optically thick models have $H/R < 10^{-3}$.  Predictions for $T_{low}$ and $N$ from the optically thin models (assuming a slab thickness limit of $H/R \ge 0.01$) are listed in Table \ref{table}. We do not consider the optically thick models realistic as they would require log($N_{tot})>51$ and imply an enormous gas:dust ratio and hydrogen abundance, but this scenario cannot be ruled out with the current data set.

\section{Discussion}
\label{discussion}
The primary pathway for OH production considered in models of irradiated environments involves the recombination of O and H atoms after destruction of H$_2$ and CO by UV radiation \citep{stoerzer98} or collisional processes \citep{kamp04}. Excited OH can also be formed directly from the photolysis of H$_2$O, producing emission known as ``prompt emission" \citep{carrington64, mumma71,bonev06}. However, in comets the OH responsible for prompt emission is not characterized by a single rotational temperature, so it is most likely not responsible for the emission reported in this paper. Understanding the role of these OH formation and excitation processes and where they occur in circumstellar disks has the potential to provide a number of constraints on disk structure and chemistry.

Th estimates for our derived inner radii using the most recent interferometry results ($R_{in} \sim1$ AU) and our derived temperatures (T $\sim$ 700K) suggest the detected OH emission originates in the surface layers of the disk.  The line shapes for OH are equivalent within error to those found for low-J CO $v= 1-0$ ro-vibrational transitions \citep{blake04}, suggesting a common inner radius.

Our results for rotational temperature and OH abundance can also be compared with recent chemical models of circumstellar disks. De-coupled gas-dust models by \citet{kamp05} predict high disk temperatures ($T > 500$K) just above the optically thick region of the disk midplane owing to the confluence of radiative and viscous heating, resulting in the production of a population of warm OH in the disk atmosphere; below the optically thick dust boundary OH production ceases. For AB Aurigae, this model predicts vertical column densities of $N_{col,v} \sim10^{13}$ \col\ for the warm OH, extending from the inner disk out to 15 AU (I. Kamp, private communication). For a 1 \Msun\ star, \citet{markwick02} also predict OH column densities of $10^{12}-10^{14}$ \col\ at 10 AU; however, the location and temperature of the OH produced in these models is uncertain. If we assume an outer radius of 15 AU and a constant vertical column density, results from our fluorescence model imply $N_{col,v} \sim 10^{13}$ \col\ for AB Aurigae and $N_{col,v} \sim 10^{14}$ \col\ for MWC 758. Results from our LTE model would be two orders of magnitude larger than the model predictions for AB Aurigae, but are less divergent for MWC 758. 

In concert with this study, \citet{salyk08?} reported a detection of OH and H$_2$O in two T Tauri stars, with inner radii and temperatures similar to those found here.  They use optically thick CO lines to find an emitting area ($0.2 - 0.4$ AU$^2$); converting their column densities to total molecules gives log$(N_{tot}) \sim 43$ based on an LTE excitation model. These abundances are consistent with ours; additional analysis of the two samples will be presented in future work. Sampling the same spectral region, we do not detect H$_2$O emission lines in either AB Aurigae or MWC 758.  Using the 3-$\sigma$ uncertainty for the spectral region as an upper limit for a line detection and applying the LTE analysis from above, we derive preliminary upper limits on the abundance of emitting H$_2$O that are lower than our derived OH abundance (see Table \ref{table}).  HAe stars have high UV fluxes that may dissociate H$_2$O in the disk surface, but a detailed model of the excitation and dissociation physics of water under these conditions has not yet been constructed.

\acknowledgements
This research was supported by the Goddard Center for Astrobiology and the NASA Post-doctoral Fellowship Program. We thank the anonymous referee for constructive comments.

{\it Facilities:} \facility{Keck:II (NIRSPEC)}

\bibliographystyle{apj}

\clearpage

\tabletypesize{\scriptsize}
\begin{deluxetable}{llccc}
\tablewidth{0pt}
\tablecaption{Quantities Derived in This Work}
\tablecolumns{4}
\renewcommand{\arraystretch}{1.4}
\tablehead{
\colhead{Model} & \colhead{} & \colhead{} & \colhead{AB Aurigae} & \colhead{MWC 758}
}
\startdata
\multicolumn{2}{l}{Inner OH Radius\tablenotemark{a}}   & E04 & $0.17^{+0.84}_{-0.17}$ AU & $1.29^{+0.59}_{-0.44}$ AU \\[1.2ex]
                                 &                                                                & I06 & $1.05^{+0.52}_{-0.30}$ AU & $1.62^{+0.26}_{-0.21}$ AU \\[1.0ex]
\cline{1-5} \\[-1.5ex]
LTE    &                  & T$_{ex}$             &  651K$\pm$41K      & 746K$\pm$39K\\[0.5ex]
           & OH:          &  log($N_{tot}$)   &  44.20$\pm$0.28   & 43.73$\pm$0.28 \\[0.5ex]
           & H$_2$O: &  log($N_{tot}$)   &  $< 44.4$   & $< 43.7$ \\[0.5ex]
\cline{1-5} \\[-1.5ex]
\multicolumn{2}{l}{Fluorescence}     & T$_{v=0}$     &  690K$-$725K       & 751K$-$808K \\[0.5ex]
                       &  OH:                            &  log($N_{tot}$)  &  42.0$-$42.9  & 43.5$-$44.5 
\enddata
\tablenotetext{a}{Calculated using disk inclinations from \citet{eisner04}(E04) and \citet{isella06}(I06).}
\label{table}
\end{deluxetable}
\tabletypesize{\normalsize}

\clearpage

\end{document}